\documentclass[prl,aps,twocolumn,superscriptaddress,showpacs]{revtex4}

\pdfoutput=1

\usepackage{amsmath}
\usepackage{amssymb}
\usepackage{graphicx}
\usepackage{float}
\usepackage{bbm}
\usepackage{mathptm}

\newcommand{\ket}[1]{\left| #1 \right\rangle}

\begin{document}

\title{Andreev-like reflections with cold atoms}

\author{A. J. Daley}
\affiliation{Institute for Quantum Optics and Quantum Information of the
Austrian Academy of Sciences, A-6020 Innsbruck, Austria} \affiliation{Institute
for Theoretical Physics, University of Innsbruck, A-6020 Innsbruck, Austria}
\author{P. Zoller}
\affiliation{Institute for Quantum Optics and Quantum Information of the
Austrian Academy of Sciences, A-6020 Innsbruck, Austria} \affiliation{Institute
for Theoretical Physics, University of Innsbruck, A-6020 Innsbruck, Austria}
\author{B. Trauzettel}
\affiliation{Department of Physics and Astronomy, University of Basel, Basel, Switzerland}
\affiliation{Institute of Theoretical Physics and Astrophysics, University of
W{\"u}rzburg, D-97074 W{\"u}rzburg, Germany}
\date{October 14, 2007}

\begin{abstract}
We propose a setup in which Andreev-like reflections predicted for 1D transport systems could be observed time-dependently using cold atoms in a 1D optical lattice. Using time-dependent Density Matrix Renormalisation Group methods we analyse the wavepacket dynamics as a density excitation propagates across a boundary in the interaction strength. These phenomena exhibit good correspondence with predictions from Luttinger liquid models and could be observed in current experiments in the context of the Bose-Hubbard model.  
\end{abstract}
 \maketitle

The rich physics described by Luttinger liquid (LL) theory \cite{Giama04} is normally associated with interacting one-dimensional (1D) electron systems such as carbon nanotubes or lithographically defined quantum wires. However, exciting progress in cold atomic gases experiments \cite{1dolexp,1dexp} has seen aspects of this physics realised in a new context \cite{1dtheory,spincharge}. This not only promises observation of effects such as spin-charge separation \cite{spincharge} in a clean system closely realising the theoretical models, but also provides a new viewpoint on transport properties, which can be studied in the context of coherent wavepacket propagation. This connection is strengthened by the use of recently developed time-dependent density matrix renormalisation group (t-DMRG) methods \cite{vidal}, which allow the computation of dynamics for physically realisable lattice models, and the identification of parameter ranges in which LL model predictions can be observed in experiments. Here we investigate this analogy for systems described by the inhomogeneous LL model, in which the electron-electron interaction varies stepwise from essentially non-interacting to repulsively interacting \cite{Maslo95}. This model is used to describe the coupling of quantum wires to higher dimensional leads which act as weakly interacting electron reservoirs, and predicts rich boundary phenomena, including Andreev-like reflection, i.e., reflection of hole excitations. We show that analogous Andreev-like reflections can exist for 1D atomic gases in optical lattices, where the many-body dynamics are well described by Hubbard models \cite{opticallattices1}, and that these could be observed time-dependently in current experimental setups. 

\begin{figure}[tb]
    \begin{center}
       \includegraphics[width=8.5cm]{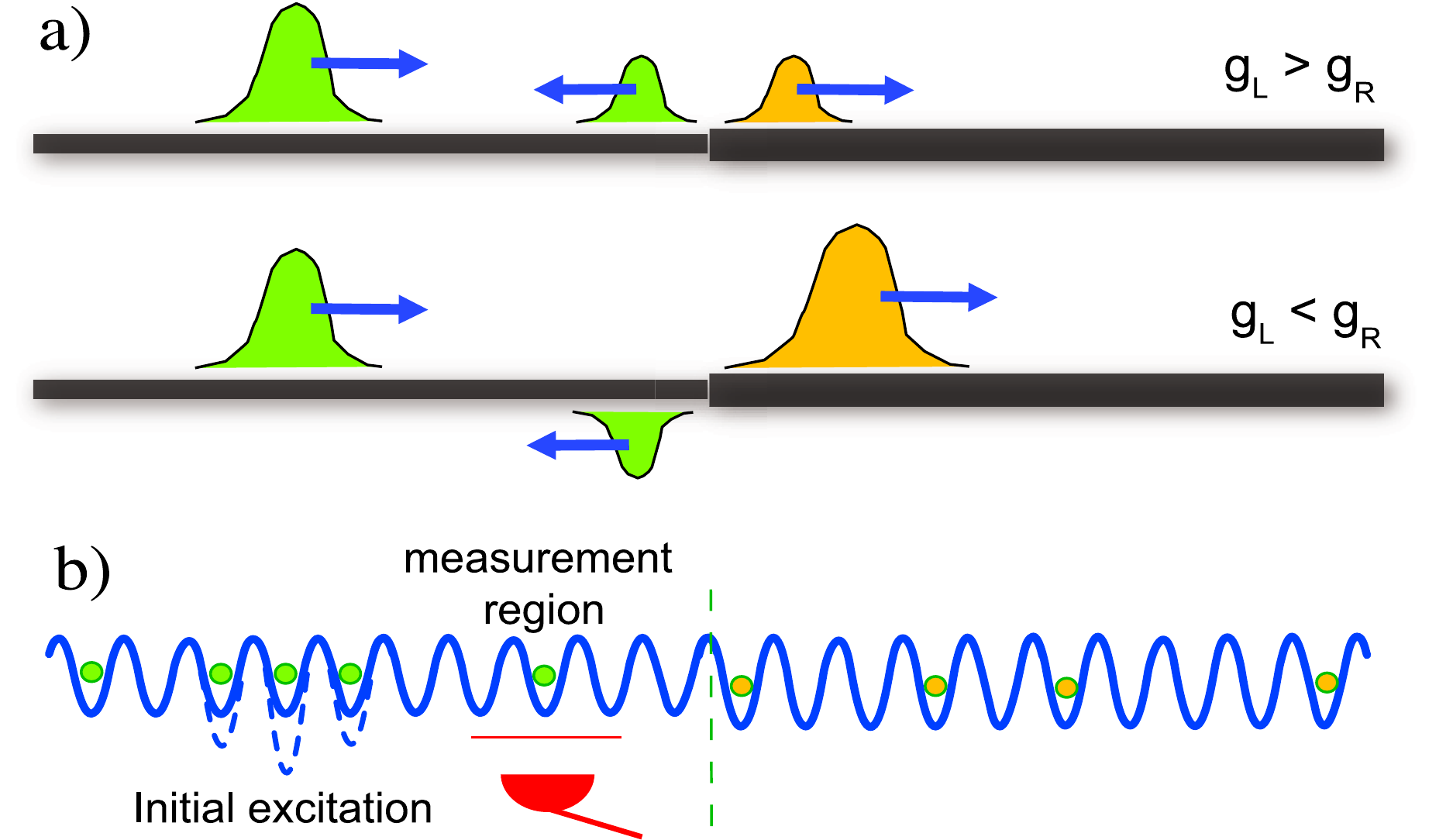}
       \caption{(a) A propagating excitation exhibits normal reflections (top) or Andreev Reflections (bottom) at an interaction boundary depending on the relative interaction strengths on the two sides. (b) Observation via bosons in an optical lattice in 3 steps: Preparation of the initial excitation using a superimposed trap (left); propagation of the excitation towards the interaction boundary, formed by coupling the atoms off-resonantly to an additional internal state (right); and detection via measurement of the atom density in a region between the location of the initial excitation and the interaction boundary.}\label{Fig:Setup} \label{fig1}
    \end{center}
\end{figure}

Andreev-like reflections are predicted by an inhomogeneous LL model  with Hamiltonian \cite{Maslo95} ($\hbar=1$) $H_{LL}=\int (dx/2\pi)[u(x) g(x) (\pi\Pi)^2+u(x)(\partial_x\Phi)^2/g(x)]$, where $\Phi$ is the standard Bose field operator in bosonization
\cite{Giama04}, and $\Pi$ its conjugate momentum density, $[\Pi(x),\phi(x')]=i\delta(x-x')$. This model describes low energy excitations with speed of sound $u(x)$. The parameter $g(x)$ characterises interactions, with $0<g(x)<1$ for repulsive interactions, $g(x)=1$ in the non-interacting case, and $g(x)>1$ for attractive interactions. When a propagating electron-like charge excitation (or density excitation) is incident on a boundary with $g(x)=g_L$ on the left of the interaction boundary and $g(x)=g_R$ on the right of the boundary, the strength of the reflections are quantified by a reflection coefficient $\gamma = (g_L-g_R)/(g_L+g_R)$. For $\gamma>0$, excitations are partly reflected and partly transmitted at the boundary (see Fig.~\ref{fig1}a, top). However if $\gamma<0$ then excitations are transmitted with a larger amplitude $1-\gamma$ which is compensated by the reflection of hole-like charge excitations with amplitude $|\gamma|$ (see Fig.~\ref{fig1}a, bottom). This is analogous to Andreev reflection when an electron is incident on a normal metal-superconductor boundary: The electron forms a Cooper pair in the superconductor, and depending on whether its energy is higher or lower than the superconducting gap, a partial or complete hole is reflected. Here there is no gap, and so the holes reflected are partial holes \cite{Maslo95}. This phenomenon is manifest in several effects predicted for transport through quantum wires -- such as oscillations of
the nonlinear current voltage characteristics and the appearance of fractional
charge excitations in the finite frequency current noise \cite{Dolci05}. However, imperfections including contact resistance between quantum wires and the attached electron reservoirs have so far prevented these effects from being observed. In this sense, cold atoms in optical lattices would constitute an ideal physical system in which Andreev-like reflections can be observed.

We study dynamics on the lattice because the physics of atoms in optical lattices is well understood on a microscopic level \cite{opticallattices1}, and t-DMRG methods allow exact computation of the dynamics. We first investigate an extended Hubbard model with offsite interactions for spin-polarised fermions (or hard-core bosons), which corresponds in the continuum limit to a Luttinger liquid \cite{Giama04}. The Hamiltonian is given by ($\hbar=1$)
\begin{equation}
\hat{H}= -J \sum_{\langle i,j\rangle }\hat c_i^\dag \hat c_j + \sum_i V_i \hat{n}_i \hat{n}_{i+1} +\sum_i \varepsilon_i \hat n_i, \label{eq:exhubbardham}
\end{equation}
where $\hat c_i$ annihilates a fermion (or boson) on site $i$, $J$ is the tunnelling rate between neighbouring sites, $n_i=\hat c_i^\dag \hat c_i$ is the number operator for particles on site $i$, $V_i$ is the nearest neighbour interaction energy and $\varepsilon_i$ denotes the energy offset of site $i$ due to external potentials. This Hamiltonian is valid for $J,V_i\overline{n} \ll  \omega$, with $\overline{n}$ the mean density, and $\omega$ the band separation. In the limit $a V_i/v_F\gtrsim 1$, where $v_F$ is the Fermi velocity and $a$ the lattice spacing, the connection between LL physics and this model is approximately given by  $g_i=1/\sqrt{1+a V_i/v_F}$. Outside of this limit, a connection can be drawn by fitting the Luttinger form for density-density correlations to numerically calculated results for the lattice model.

Offsite interactions can be generated with Fermions, e.g., using polar molecules or by coupling to Rydberg states, or with hard-core Bosons by loading strongly interacting atoms into excited Bloch bands \cite{offsiteint}. However, the natural experimental situation is to have short range contact interactions between atoms, as described by the Bose-Hubbard model including only on-site interactions, with Hamiltonian ($\hbar=1$)
\begin{equation}
\hat{H}= -J \sum_{\langle i,j\rangle }\hat b_i^\dag \hat b_j + \sum_i U_i \hat{n}_i (\hat{n}_{i}-1) +\sum_i \varepsilon_i \hat n_i. \label{eq:bhham}
\end{equation}
Analogously to eq. (\ref{eq:exhubbardham}), $\hat b_i$ annihilates a boson on site $i$, $J$ is the tunnelling rate between neighbouring sites, $n_i=\hat b_i^\dag \hat b_i$ is the number operator for particles on site $i$, $U_i$ denotes the onsite interaction energy shift between two atoms and $\varepsilon_i$ denotes the energy offset of site $i$ due to external potentials. This Hamiltonian is valid in the limit where $J, U_i\overline{n} \ll \omega$. Note that in the limit $|U_i/J|, |U_i/\varepsilon_i| \gg 1$ and with a mean filling factor $\bar n \ll 1$, we can obtain the off-site interactions of the extended Hubbard Hamiltonian, eq. (\ref{eq:exhubbardham}) directly from onsite interactions. Restricting to the manifold of states containing only singly occupied sites, we obtain off-site interactions in perturbation theory as $V_{i, \rm eff} =-J^2/U_i-J^2/U_{i+1}$. Below we also go beyond this limit in our numerical calculations.

We begin by studying dynamics in the extended Hubbard model, eq.~(\ref{eq:exhubbardham}) before returning to the Bose-Hubbard model below. In analogy to the case in quantum wires, we would like to observe a density excitation propagating through the system towards a boundary in the interaction strength. This density excitation is formed by creating a local dip in the external potential (e.g., using a focussed laser beam, see Fig.~\ref{fig1}b),
\begin{equation}
\varepsilon_i=- \varepsilon_0 \exp [-(i-x_0)^2/(2 \sigma^2)] + \varepsilon_R  F (i).
\end{equation}
Here, $\sigma$ denotes the desired Gaussian width of the local density excitation, centred on an initial site, $x_0$, and $\varepsilon_0$ will control the depth of the potential and hence the maximum density of the excitation. We choose the state for $t<0$ to be the ground state of the Hamiltonian (\ref{eq:exhubbardham}), with $\varepsilon_0(t<0)>0$. At $t=0$, we switch off the local dip in the external potential suddently, $\varepsilon_0(t)=\varepsilon_0 \theta (-t)$, leaving a Gaussian shaped density excitation.
The last term denotes a difference in potential between the left and right of the barrier, with the barrier function $F(x)$, which beginning at site $x_b$ will be taken to vary linearly across the width of the barrier, $M_b$ sites, as $F(i)=0,$ $i<x_b$; $F(i)=(i-x_b)/M_b$, $x_b\leq i\leq x_b+M_b$; $F(i)=1$, $i>x_b+M_b$. The interaction $V_i$ will vary in the same way as $V_i=V_L + (V_R-V_L) F(i)$. Initially we will consider the boundary to be sharp, i.e., there is a step at a given lattice site $x_b$, so that $V_i=V_L$ for $i<x_b$, and $V_i=V_R, i\geq x_b$. The parameter $\varepsilon_R$ should be adjusted so that the density on each side of the barrier is approximately the same. Note that obtaining a constant initial density close to the barrier is only possible in the range $|V_L-V_R| \lesssim J$, as otherwise large oscillations are observed in the density near the boundary. 

The initial groundstate (with $\varepsilon_0>0$) and subsequent propagation are computed by quasi-exact imaginary and real time evolution respectively under the Hamiltonian (\ref{eq:exhubbardham}). This is made possible by t-DMRG methods \cite{vidal}, which are applicable to 1D many-body systems where the Hilbert space can be expressed as the product of a chain of local Hilbert spaces. The state of the system is effectively written as a truncated Matrix Product state representation \cite{vidal}, in which $\chi$
states are retained in each Schmidt decomposition of the system. In our calculations we performed convergence tests to ensure the accuracy of our results (see \cite{Gobert} for a general analysis of the accuracy of these methods), and estimate errors smaller than a few percent in the presented values.

\begin{figure}[tb] 
    \begin{center}
       \includegraphics[width=8.5cm]{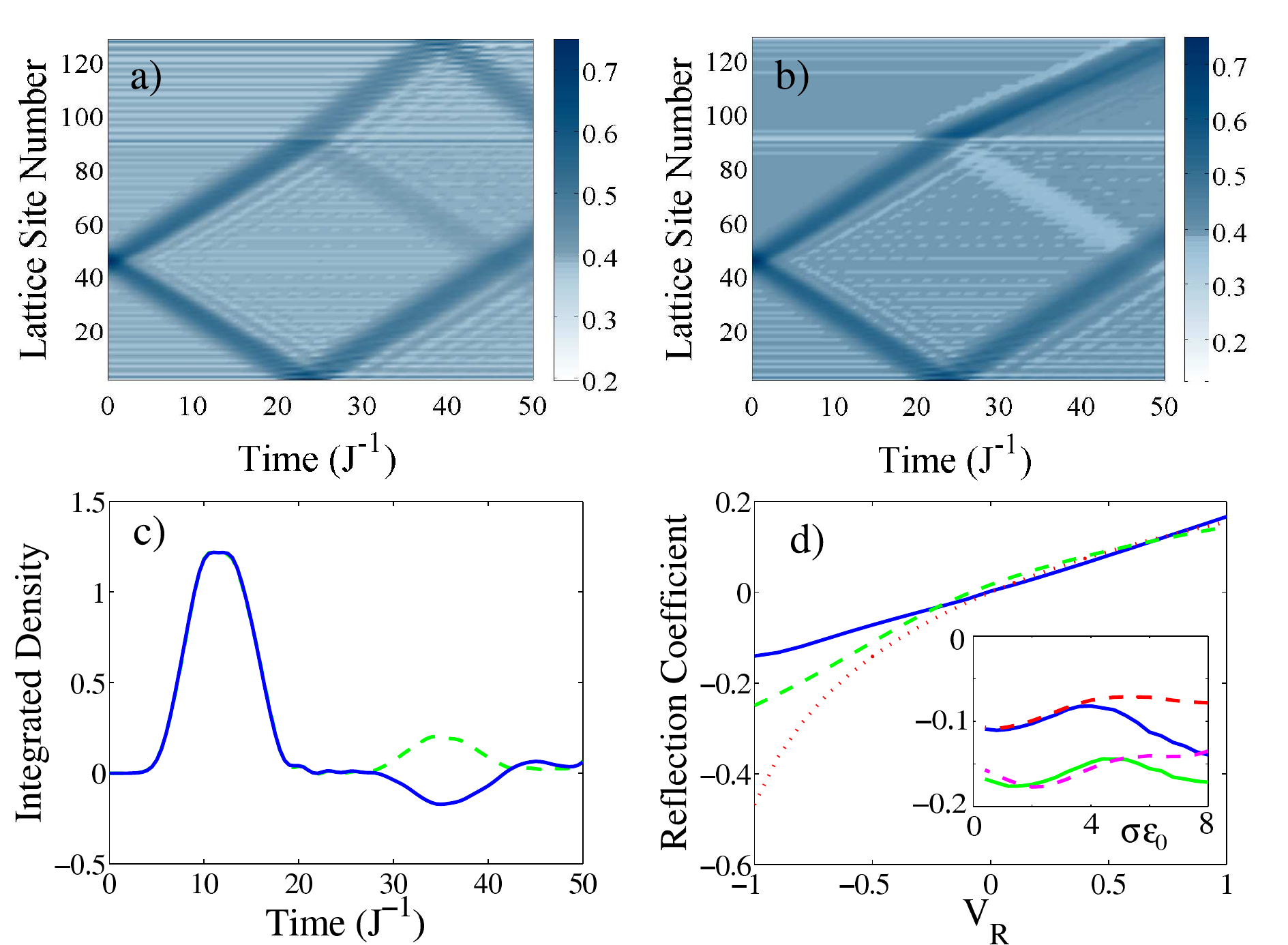}
       \caption{Numerical simulation results for propagation of an initial density excitation ($\varepsilon_0=2J$, $\sigma=3$, $x_0=45$) across an interaction boundary from $V_L=0$ to varying $V_R$ at $x_b=90$ ($M_b=1$), with open boundary conditions. (a) Shaded density plot for $V_R=J$, showing a normal (positive density) reflection at the boundary. (b) Shaded density plot for $V_R=-J$, showing an Andreev (negative-density, or hole) reflection at the boundary.  (c) Difference from the initial value of the integrated density over sites 60--75, showing initial peak due to propagation of the initial density excitation, and a secondary maximum ($V_R=J$, dashed line) or minimum ($V_R=-J$, solid line) due to the reflected excitation. (d) Reflection coefficients as a function of $V_R$, showing the comparison between simulation values (solid line), and estimated parameters from LL theory using numerically computed Luttinger parameters (from eq.~(\ref{eq:densdens}), dashed line) and the analytical form $g=1/\sqrt{1+a V_i/v_F}$ (dotted line). The inset shows the dependence on the depth $\epsilon_0$ with $\sigma=2$ (solid lines) and $\sigma=4$ (dashed lines) for $V_R=-0.5 J$ (upper curves) and $V_R=-J$ (lower curves), computed from measurement sites 50--70.}\label{fig2}
    \end{center}
\end{figure}

In Fig.~\ref{fig2}a,b we show shaded plots of the density of atoms at different lattice sites as a function of time. From these plots we clearly see the propagation of the initial density excitation, which splits into a right-moving and left-moving excitation. The left-moving excitation is incident on the left-hand boundary of the complete system, and plays no role in the following discussion. The right-moving excitation, however, is incident on the interaction boundary, resulting in reflected as well as transmitted excitations. In Fig.~\ref{fig2}a, where $V_L=0$ and $V_R=J$, we see a normal reflection of the excitation, and both the transmitted and reflected excitations have a smaller density than the incident excitation. In Fig.~\ref{fig2}b, we have $V_R=-J$, and we observe an Andreev reflection, where a hole excitation is reflected, corresponding to a lower density at the sites it occupies. Commensurately, a larger amplitude excitation is transmitted at the interaction boundary. 

We can quantify this process by defining the amplitude of a density excitation to be the total integrated change from the background density over the sites containing the excitation. The reflection coefficient $R$ naturally follows as the ratio of amplitudes of the reflected and incident excitations. In practice, we identify a group of sites between the original location of the density excitation and the interaction boundary as the measurement region. As shown in Fig.~\ref{fig2}c, the density in the measurement region increases and then decreases as the initial right-moving excitation passes, then we observe either an increase or decrease in the density resulting from normal or Andreev-like reflections respectively. If the measurement region contains some part of the initial excitation, then the background density is computed between the time windows where the initial and reflected excitations pass. The reflection coefficient can be found as the ratio of the peak values in each of these time periods. Note that this definition can be used operationally in an experiment, where the integrated density over several sites can be measured, e.g., using flouresence or phase-contrast imaging with a focussed laser (see Fig.~\ref{fig1}b).

In Fig.~\ref{fig2}d we plot the reflection coefficient $R$ for $-J<V_R<J$, choosing $V_L=0$, $\varepsilon_0=2J$, and $\sigma=3$, and see clearly the crossover from Andreev to normal reflection. Note that the quantitative values (though not the behaviour with varying $V_R$) are dependent on the size of the initial excitation, as shown in the inset of Fig.~\ref{fig2}d. In order to compare our results with the known analytical result from LL theory, we extracted approximate Luttinger parameters $g_{\rm eff}$ for the ground states of our extended Hubbard model, eq. (\ref{eq:exhubbardham}), at the background density value for each $V_R$. This was done by computing density-density correlations, and fitting the standard form for these correlations in a Luttinger liqud \cite{Giama04},
\begin{equation}
\langle \hat n_0 \hat n_r \rangle \approx  -\frac{g_{\rm eff}}{2\pi^2 r^2}+ A\cos(2\pi \bar n r)\left(\frac{1}{r}\right)^{2g_{\rm eff}}, \label{eq:densdens}
\end{equation}
where $A$ is a constant, and $\bar n$ is the mean occupation per lattice site.
Using these values we computed approximate reflection coefficients, which are plotted in Fig.~\ref{fig2}d. They show very good quantitative agreement with $R$ in our simulations for $V_R>0$, and only small deviations for $V_R<0$, especially for initial density excitations containing approximately one particle. This agreement is better than might be expected, and demonstrates both the generality of LL theory in its applicability to low energy states of 1D systems, and its continued applicability when we introduce by hand the boundary in the interaction strength. For comparison we have also computed $g_{\rm eff}$ based on the approximate analytical expression.

\begin{figure}[tb]
    \begin{center}
        \includegraphics[width=8.5cm]{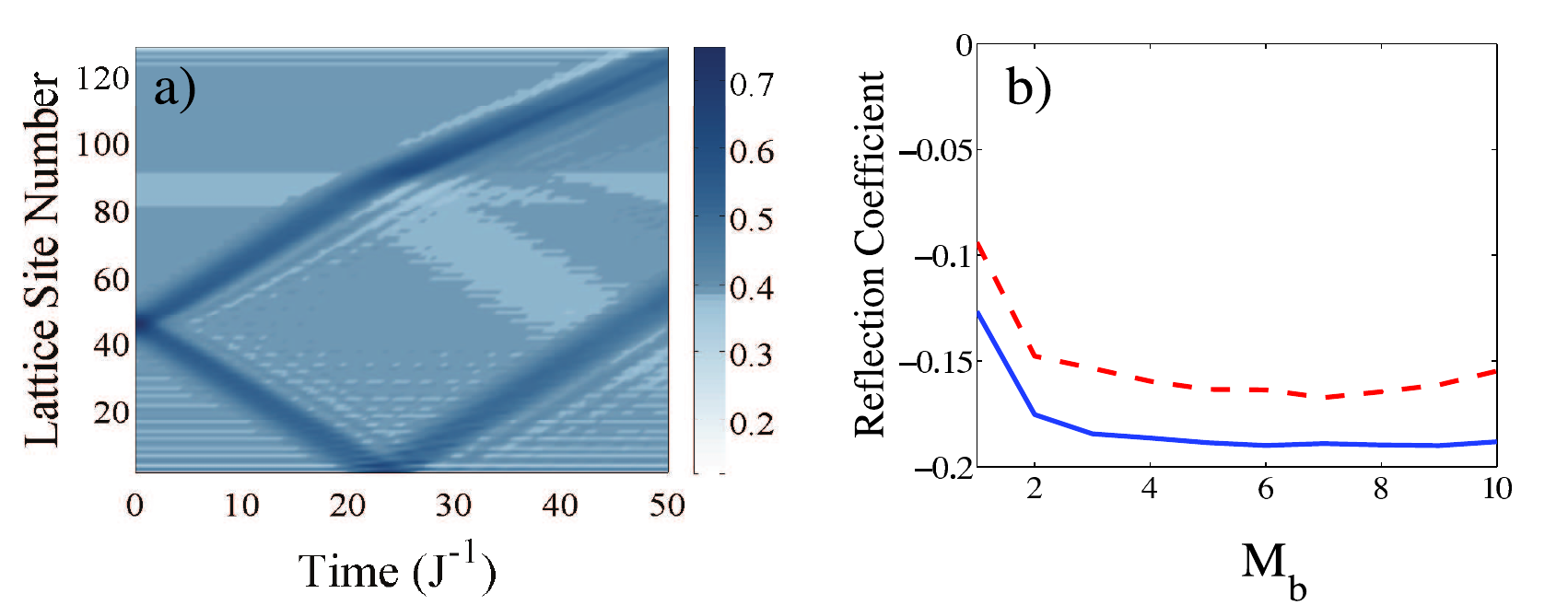}
       \caption{Reflections of a density excitation ($\varepsilon_0=2J$, $x_0=45$) from a thick boundary with the interaction strength varying from  $V_L=0$ to $V_R=-J$ linearly over $M_b$ lattice sites. (a) Shaded density plot showing reflection of an initial density excitation with $\sigma=3$ from a boundary with $M_b=10$, showing the considerably broader reflected wave produced by the thick boundary. (b) Reflection coefficient as a function of barrier thickness $M_b$ (measurement sites 55--85), showing an increase in the amplitude of the negative density reflection as the barrier is increased, for initial wavepackets with $\sigma=3$ (solid line) and $\sigma=2$ (dashed line). The amplitude is also slightly larger for the narrower initial wavepacket. $\varepsilon_0(t=0)=2J$.}\label{Fig:Res2}\label{fig3}
    \end{center}
\end{figure}

\begin{figure}[tb]
    \begin{center}
        \includegraphics[width=8.5cm]{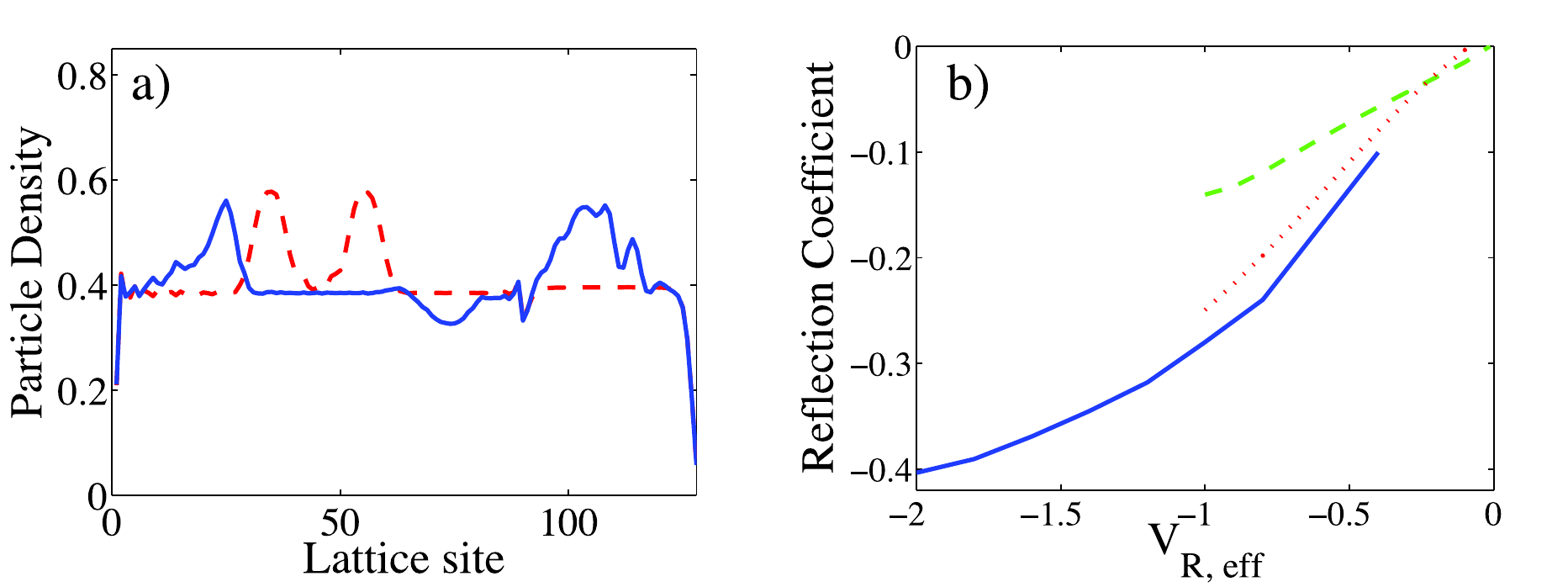}
       \caption{Realisation of Andreev reflections within the Bose-Hubbard Model: (a) The density $\langle \hat n_i \rangle$ at each lattice site at time $tJ=5$ (dashed line)  showing the two counterpropagating density excitations, at time and $tJ=30$ (solid line), showing the transmitted and reflected excitation with $U_R=J$. Note the clear negative-density excitation centred near site 70. (d) Reflection coefficients for varying $U_R$, plotted as a function of $V_{R,{\rm eff}}$ for the Bose-Hubbard model (solid line). These are compared with results from the extended Hubbard model (dashed line) and estimated parameters from LL theory (dotted line) computed as in Fig.~2. Parameters used were $\varepsilon_0=2J$, $\sigma=3$, $x_0=45$, $M_b=1$, $U_L=10J$, measurement sites 55--85.}\label{Fig:ol}\label{fig4}
    \end{center}
\end{figure}

Until now we have considered only sharp barriers with an immediate transition from $V_L$ to $V_R$. We now investigate the time propagation of the excitations when $V_i$ varies linearly over $M_b$ sites, which provides a more realistic treatment of barriers that might be created in a real experiment. An example shaded plot of the density of atoms at different lattice sites as a function of time for an excitation exhibiting Andreev reflection with $V_L=0,\, V_R=-J$ is shown in Fig.~\ref{fig3}a. We see clearly that the extended length barrier spreads the resulting reflected and transmitted excitations in space (cf. Fig.~\ref{fig2}b). However, the total amplitude of the reflected wavepacket is actually increased in this case. This can be seen from Fig.~\ref{fig3}b, where we plot the reflection coefficient $R$ as a function of $M_b$ for different amplitudes of the incident excitation. This brings the reflection coefficients closer to the values obtained from the effective LL parameters. The finite-width barrier also smoothes the local density minima or maxima that appear at the boundary in the ground state (see Fig.~\ref{fig4}a).

To provide a simpler experimental implementation, we consider the dynamics of excitations at an interaction boundary in the Bose-Hubbard model. In the perturbation theory limit, it follows from $|J/U|\ll 1$, that $|V_{i, \rm eff}| \ll J$, and thus the amplitude of all reflections will be extremely small. However, using numerical simulations we can treat this system exactly also beyond the limit in which perturbation theory is valid. In Fig.~\ref{fig4}a we show results from numerical simulations in which a density excitation (created as for the extended Hubbard model) propagates across a interaction boundary at $x_b$ with $U_{i<x_b}=10J$ and $U_{i\geq x_b}=J$. We see very clearly that a hole excitation is reflected, and reflection coefficients as a function of $V_{L, \rm eff}=1/U_{i>x_b}$ are plotted in Fig.~\ref{fig4}b., and compared with results for the extended Hubbard model with $V_{i>x_b}=V_{i, \rm eff}$. Effects that go beyond the validity of perturbation theory actually yield a slight increase in the amplitude of Andreev-like reflections, and bring these results close to those predicted using the estimated Luttinger parameters that were plotted in Fig.~\ref{fig2}d. In addition, we can use values of $U\sim J$, where $|V_{\rm eff}/J|>1$  without adverse boundary effects that prevent us from obtaining a smooth background density near the interaction boundary. Thus, it is possible to observe even larger amplitude Andreev-like reflections in the Bose-Hubbard model than in the extended Hubbard model. 

In an experiment, the change in onsite interaction strength could be engineered simply in several ways. For example, lasers focussed on one side of the system could couple the atoms off-resonantly from their internal state, $\ket \alpha$ to an additional internal state $\ket \beta$. If a Feshbach resonance \cite{feshbach} exists between two atoms both in state $\alpha$ so that their interaction $U_{\alpha \alpha}$ is much larger than that between atoms in the two different states $U_{\alpha \beta}$, then adding an admixture of the state $\beta$ will reduce the onsite interaction strength in the region where the internal state of the atoms is $\ket\psi=a_1 \ket\alpha + a_2 \ket\beta$, where $a_1$ and $a_2$ are complex coefficients. Such a laser coupling could reasonably be focussed so that the coupling varies on a length scale of $\lesssim 5 \mu$m, or approximately 10 lattice sites. Similarly, the initial density excitation could be prepared using a laser focussed over $\sim 10$ lattice sites.

We have shown how time-dependent wavepacket dynamics of Andreev reflections in Hubbard models closely match the behaviour expected from LL physics, and how these reflections could be observed time-dependently with cold atoms in optical lattices. This work could be extended to multi-species models, such as fermionic Hubbard models, and to other geometries such as Y-junctions \cite{andreevdemler}. The boundary physics in each case has different characteristics which could be explored time-dependently in the experiments.

We thank U. Schollw\"ock, A. Kantian, N. Davidson, A. M. Rey and G. Pupillo for interesting discussions. Work in Innsbruck was supported by the Austrian FWF through project I118\_N16 (EuroQUAM\_DQS) and SFB F15, and by the EU networks OLAQUI and SCALA. Work in Basel was supported by by the Swiss NSF and the NCCR Nanoscience.

\end{document}